\documentclass[pra,showpacs,twocolumn,showkeys]{revtex4-1}
\usepackage{graphics}
\usepackage{graphicx}
\usepackage{amsmath}
\newcommand{\bibs}{C:/Users/Xavi/Dropbox/References/BibFile}

\begin{document}

\title{Applying universal scaling laws to identify the best molecular design paradigms for second-order nonlinear optics}

\author{Javier Perez-Moreno}
\email{jperezmo@skidmore.edu} \affiliation{Department of Physics, Skidmore College, Saratoga Springs, New York 12866}

\affiliation{Department of Physics and Astronomy, Washington State University, Pullman, Washington 99164-2814}

\author{Shoresh Shafei}
\affiliation{Department of Physics and Astronomy, Washington State University, Pullman, Washington 99164-2814}

\affiliation{Current address: Department of Chemistry, Duke University, Durham, North Carolina 27708-0354}

\author{Mark G. Kuzyk}
\affiliation{Department of Physics and Astronomy, Washington State University, Pullman, Washington 99164-2814}

\begin{abstract}
We apply scaling and the theory of the fundamental limits of the second-order molecular susceptibility to identify material classes with ultralarge nonlinear-optical response.  Size effects are removed by normalizing all nonlinearities to get intrinsic values so that the scaling behavior of a series of molecular homologues can be determined.  Several new figures of merit are proposed that quantify the desirable properties for molecules that can be designed by adding a sequence of repeat units, and used in the assessment of the data.  Three molecular classes are found.  They are characterized by sub-scaling, nominal scaling, or super-scaling.  Super-scaling homologues most efficiently take advantage of increased size.  We apply our approach to data currently available in the literature to identify the best super-scaling molecular paradigms with the aim of identifying desirable traits of new materials.
\end{abstract}

\pacs{42.65.An, 33.15.Kr, 11.55.Hx, 32.70.Cs}

\maketitle

\section{Introduction}

The quest for making materials with ever-larger nonlinear-optical response is fueled by the promise of next generation devices of diverse applications such as electro-optic circuit imaging, phase and intensity modulators, harmonic generators, entangled photon generation, and second-harmonic bio-imaging.  Larger nonlinearities are being reported each day, making new applications possible. For example, Jiang et al. have reported on coupled porphyrins with a huge first hyperpolarizability of $11,300 \times 10^{-30} esu$.\cite{jiang12.01} Van Cleuvenberger et al. have found record-high (intrinsic) first hyperpolarizability from disubstituted poly(phenanthrene) polymers.\cite{cleuv14.01}  Coe et al. have investigated the second-order nonlinear optical response of Helquat Dyes with large static first hyperpolarizabilities.\cite{coe2016helquat} Al-Yasari et al. have characterized donor-acceptor organo-imido polyoxometalates with static first hyperpolarizabilities that exceed those of comparable compounds.\cite{al2016donor} Can we expect continued improvements?  How do we know which molecular paradigms are best for future improvements? How can we compare the performance of molecules of different size?

The strength of the nonlinear optical response of a molecule is limited by the number of electrons, reaching the fundamental limit only when they are optimally arranged.\cite{kuzyk00.01,kuzyk00.02,kuzyk03.02,kuzyk03.01} The performance of a molecule is evaluated by comparing its response with that of the optimal structure that uses the same number of electrons (the quantum limit).  An analysis based on the quantum limit theory applied to experimental data allows one to determine the mechanisms that dominate the nonlinear optical response at the molecular level,\cite{tripa04.01, tripa06.01, perez07.02, zhou08.01, perez05.01, perez11.02, perez11.01, perez09.02, van2012dispersion, de12.01} introduce new paradigms for optimization,\cite{perez09.01, perez07.01, perez06.01, perez11.02, Kang05.01, brown08.01, He11.01} and establish fundamental scaling laws.\cite{kuzyk10.01, kuzyk13.01} In this paper we introduce a new analysis that we apply to experimental data in the literature to identify the best molecular candidates for the largest second-order nonlinear optical response. The extension to the third-order nonlinear optical regime is straightforward and will be presented in a follow-up paper.\cite{perez16gamma}  More importantly, our approach pinpoints structural properties that lead to super scaling so that simply making the molecule larger leads to drastic enhancements.

The starting point is the identification of the simple scaling rules that determine how the size of a quantum system drives the strength of the nonlinear-optical response.\cite{kuzyk13.01}  This size dependence is called {\em simple scaling}, and must be removed to determine the intrinsic nonlinear-optical response, a measure of the molecule's intrinsic performance that is unbiased by its size.  The intrinsic nonlinear-optical response as a function of length thus yields a performance metric that places molecules into three classes: those that increase less with length than predicted by scaling -- called sub-scaling; those that scale according to predictions -- called nominal scaling; and those whose nonlinear response increases at a higher rate than predicted -- called super-scaling.

Most molecules fall into the sub-scaling class, so proposed design paradigms based on smaller molecules usually disappoint when they are made larger.  Since most molecules fall far below the fundamental limits, those that scale at or less than predictions will loose ground as they are made larger.  Though the nonlinear-optical response may be larger with size, the electrons in the larger molecules are being used less effectively, thus making them fall further short of their potential.  Molecules that have the largest intrinsic nonlinearity and super-scale have the potential for hitting the fundamental limits.   One goal of the present work is to identify molecules that super scale and are members of homologues that already have been synthesized and can be made larger to effectively take advantage of their nonlinear response.  A goal as important is the development of a technique that can be applied to both making better materials and understanding the structural properties associated with a large nonlinear-optical response.  This paper describes both.

After introducing limit theory and scaling, we propose several figures of merit that apply to a group of homologues -- which quantify the type of scaling, the extrapolated molecule size that hits the fundamental limit, and the nonlinearity at saturation.  These benchmarks are applied to many molecular classes with second-order susceptibilities, and are used to identify singular systems.

\section{Approach}

The molecular property of interest is the first hyperpolarizability, $\beta$, a third rank tensor. Since we are interested in the largest nonlinear optical response, and the largest is usually a diagonal component, we call the largest first hyperpolarizability tensor components simply $\beta$. The fundamental limit of $\beta$ is calculated using the sum rules and is given by:\cite{kuzyk00.01}
\begin{equation}\label{eq:betaMax}
\beta_{max} =	\sqrt[4]{3} \left( \frac{e\hbar}{\sqrt{m}}\right)^{2} \frac{N^{3/2}}{E_{10}^{7/2}},
\end{equation}
where $e$ and $m$ are the charge and mass of the electron, $\hbar$ is the reduced Plank constant, $N$ is the effective number of electrons, and $E_{10}$ is the energy difference between the first and the ground state.  Using $esu$ units we can express the coefficient of Equation \ref{eq:betaMax} numerically for simple evaluation, giving
\begin{equation}\label{eq:betaMaxUnits}
\beta_{max} \left[ \frac{cm^{5}}{statvolt} \right] = 1,186 \times 10^{-30} N^{3/2} /E_{10} \left[ eV \right]^{7/2},
\end{equation}
where the quantities in brackets are the units. The conversion between energy of a photon in $eV$ and its associated wavelength $\lambda$ in nanometers is $\lambda[nm] = 1240/E[eV]$.

The fundamental limit defines an absolute maximum, so the ratio of the measured nonlinearity to the limit is a dimensionless parameter of magnitude less than unity.  We define the intrinsic first hyperpolarizability as this ratio:\cite{zhou08.01, zhou07.02}
\begin{equation}\label{eq:betaInt}
\beta_{int} = \frac {\beta} {\beta_{max}}.
\end{equation}

It has been demonstrated that in general, the first hyperpolarizability scales in the same manner as the fundamental limits:\citep{kuzyk13.01}
\begin{equation}
\label{eq:betascales}
\beta \propto \frac{N^{3/2}}{E_{10}^{7/2}}.
\end{equation}
This is called {\em ``simple scaling''}. Clearly, while the absolute first hyperpolarizability follows simple scaling, the effect is eliminated by computing intrinsic quantities. Therefore, the intrinsic first hyperpolarizabilities is said to be scale invariant.

It can be shown rigorously that the Schr\"{o}dinger Equation is invariant under transformations in which the lengths are re-scaled by a factor $\epsilon$ if the energies are simultaneously re-scaled by a factor $1/\epsilon^2$.\cite{zhou08.01,kuzyk10.01} This same re-scaling leaves the intrinsic polarizability and  hyperpolarizabilities unchanged. We can expand this idea to the case of molecules and classify them in terms of their scaling behavior. We define a molecular {\em ``class''} as a collection of homologue molecules of varying sizes. If the intrinsic nonlinearities of the molecules within a class are the same (i.e. the intrinsic nonlinearity does not depend on the length), then the class obeys simple scaling. If the intrinsic nonlinearity increases with length the class is said to be super-scaling.  Finally, if the intrinsic nonlinearity decreases with length, the class obeys sub-scaling. Clearly, molecular classes with a large second-order nonlinear response that super scale are the target paradigm.  This work seeks to apply such an analysis to identify the best molecules.

\section{Results and discussions}

\begin{figure*}
\centering
  \includegraphics[width=4in]{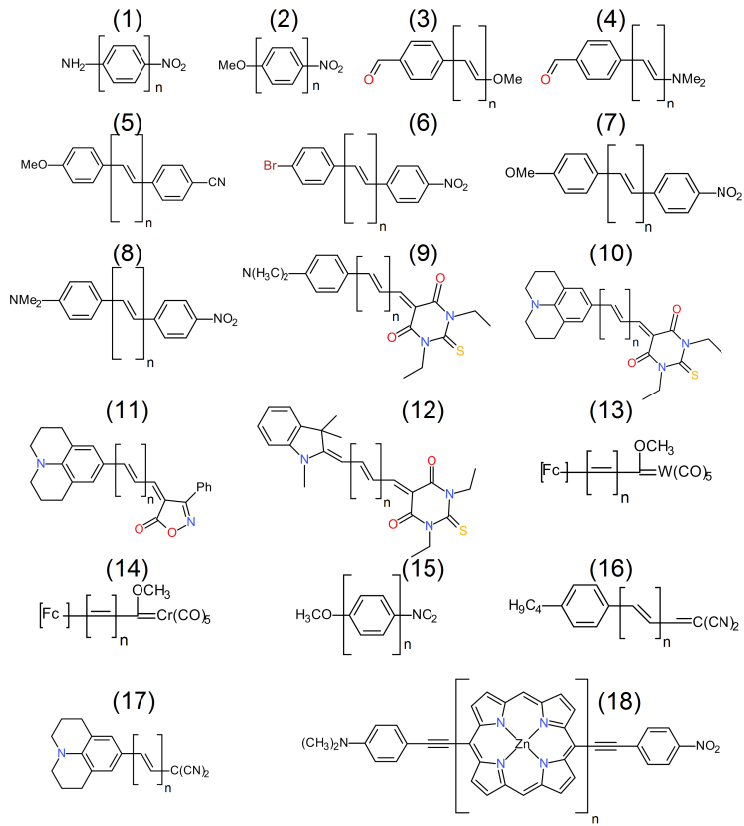}\\
  \caption{The molecular classes with measurements of the absolute first hyperpolarizability ($\beta_{exp}$) considered in this work }\label{fig:beta}
\end{figure*}

Figure \ref{fig:beta} shows the molecular classes whose first hyperpolarizabilities are studied. In each case, the base molecule is shown, and the class is defined by varying the number of repeat units, $n$. The calculation of the intrinsic first hyperpolarizability requires as an input the measured first hyperpolarizability, the effective number of electrons, $N$, and the energy difference between the ground and first electronic excited states, $E_{10}$, so that Equation \ref{eq:betaInt} can be evaluated using Equation \ref{eq:betaMax}.

The absolute first hyperpolarizabilities are determined from measurements reported in the literature, the energy differences $E_{10}$ are determined from the wavelength of maximum absorption, and the effective number of electrons are determined according to:
\begin{equation}\label{eq:betaEffN}
N_{\beta} =  \left( \sum_{i} N_i^{3/2} \right)^{2/3}.
\end{equation}
where the sum is over each contiguous conjugated path.\cite{kuzyk03.03}  For a single conjugated path, there are two electrons per double or triple bound, and the effective number of electrons is simply the total number of $\pi$-electrons in the conjugated path. The number of effective electrons is calculated using the exponent of $N$ in Equation \ref{eq:betaMax}. The values of $E_{10}$ and $N$ for all the molecular classes considered in this study are tabulated in Table \ref{tab:beta}.

\begin{figure}
  \centering
  \includegraphics{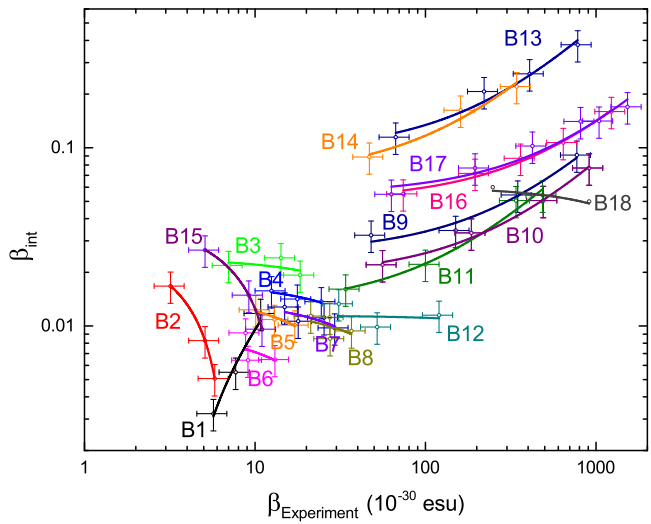}
  \caption{Plot of the intrinsic hyperpolarizability ($\beta_{int}$) as a function of the measured absolute hyperpolarizability ($\beta_{exp}$) for the 18 molecular classes considered in this study. The fit is to the function $\beta_{int} = c \beta_{exp} + d$ and the fit parameters are given in Table \ref{tab:beta}. Notice that the scales for the horizontal and vertical axis are logarithmic, and thus, the linear fits appear as curves.}\label{fig:betaIntBetaExp}
\end{figure}

Figure \ref{fig:betaIntBetaExp} shows a composite log-log plot of the intrinsic hyperpolarizability ($\beta_{int}$) as a function of the measured absolute hyperpolarizability ($\beta_{exp}$) for all molecular classes.  The data points for each molecular class are shown with a distinct color. The error bars are experimental uncertainties and the curves represent linear fits to the data of the form:
\begin{equation}
\beta_{int}(n) = c \cdot \beta_{exp}(n) + d,
\label{eq:linfitintexp}
\end{equation}
using weighted fitting, where points with smaller uncertainty carry a heavier weight.  The color of the curves are the same as the data points to which they correspond. Linear fits are chosen for simplicity and most of the data is consistent with linearity. The values of the fitting parameters for each molecular class are listed in Table \ref{tab:beta}.

Visual inspection of the plot identifies the range of $\beta_{int}$ and $\beta_{exp}$, the trends within each series of molecules, and the type of scaling.  Generally, the intrinsic nonlinear response is flat or grows with the experimental absolute hyperpolarizability.  The obvious exceptions are classes B2 and B15.  Small molecules generally have smaller absolute hyperpolarizabilities than larger ones.  For example the absolute hyperpolarizabilities of the smallest molecules are about $1 \times 10^{-30} esu$ while the largest ones are about $1,500 \times 10^{-30} esu$, a range of three orders of magnitude.  In contrast, the intrinsic hyperpolarizability varies by just over an order of magnitude.

The intrinsic hyperpolarizability removes the effects of the molecular size due to simple scaling, while the absolute hyperpolarizability is dependent on size. Therefore, it is no wonder that the range of the intrinsic nonlinear-optical response is narrower than the range of the absolute hyperpolarizability.
This narrower range of intrinsic hyperpolarizabilities illustrates how the optimization of the absolute hyperpolarizability can easily be dominated by simple scaling strategies (i.e. strategies based on elongation of the conjugated path). For example, bond length alternation,\cite{marde94.02} a tool used by many chemists as a guide to making better molecules, is a property that mostly determines the effective size of a molecule.\cite{tripa04.01}

Since the absolute hyperpolarizability depends on size as well as variations due to other factors that represent the molecule's intrinsic ability to interact with light, we remove simple scaling effects to obtain $\beta_{int}$, a direct measure of this intrinsic ability and then classify the scaling into the three types of scaling.  If $\beta_{int}$ does not change with the number of repeat units, we call it {\em ``simple scaling"}.  Otherwise the class follows {\em ``sub-scaling."} Ff $\beta_{int}$ decreases with the number of repeat units and {\em ``super-scaling"} if it increases with the number of repeat units.

\begin{figure*}
  \centering
  \includegraphics[width=5.5in]{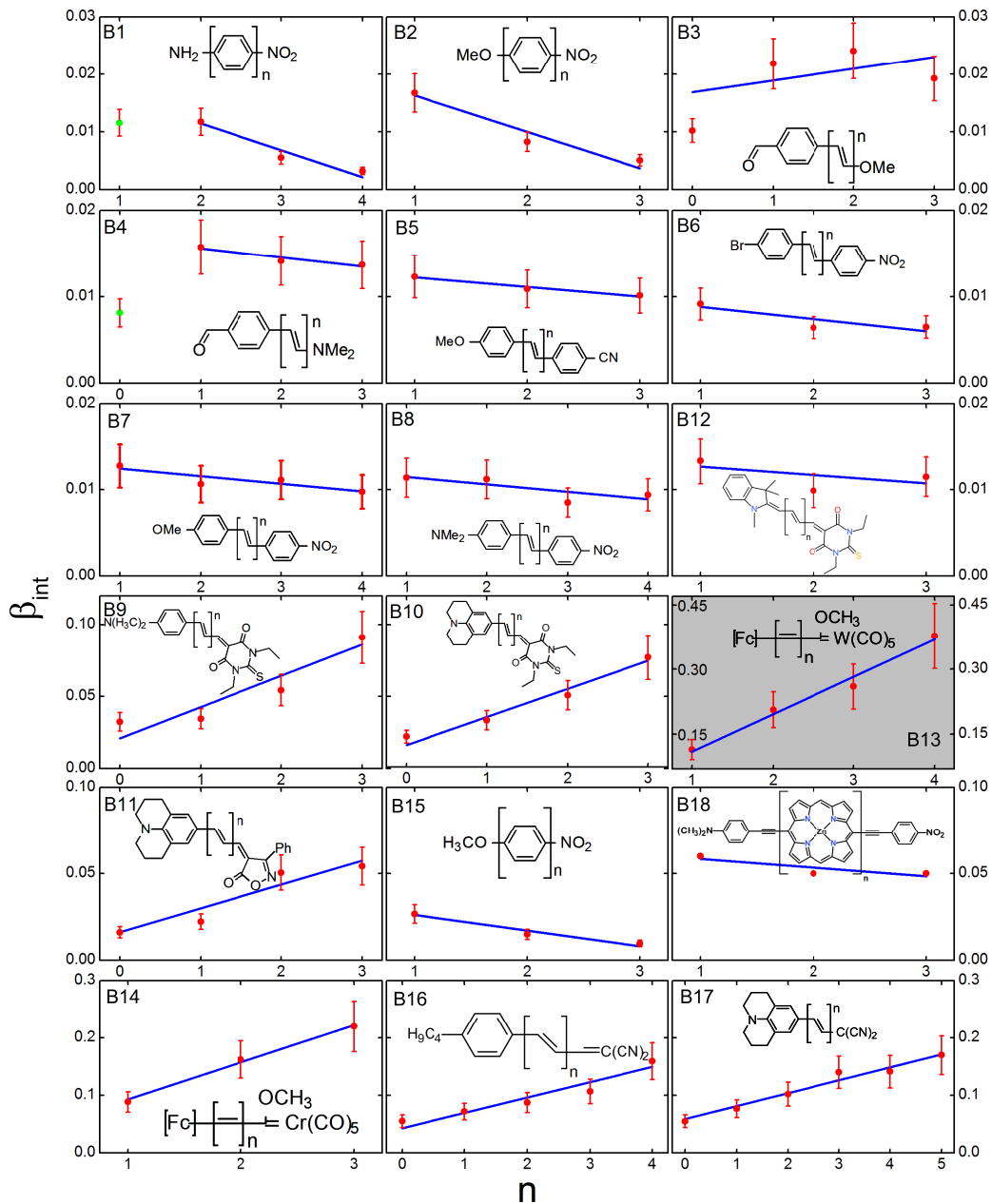}
  \caption{A plot of $\beta_{int}$ as a function of the number of repeat units, $n$.  The fit function is $\beta_{int} (n)= a \cdot n + b$ and the fit parameters are given in Table \ref{tab:beta}. The green points are outliers that have not been included in the fits. Notice that each row shares the same vertical scale for $\beta_{int}$, except for the shaded figure.}\label{fig:betaIntn}
\end{figure*}

Figure \ref{fig:betaIntn} shows plots of the experimentally-determined intrinsic hyperpolarizabilities as a function of number of repeat units (red points) and weighted linear fits (blue lines). The linear fits are of the form:
\begin{equation}
\beta_{int} (n)= a \cdot n + b.
\label{eq:linfitbetaint}
\end{equation}
The green points are outliers that have not been included in the fits. Each row shares the same $\beta_{int}$ axis except for the shaded figure, where the right axis shows its values.  The linear fits are a good approximation to the data and the pattern is similar to the one observed in Figure \ref{fig:betaIntBetaExp}, though some of the points are shifted.  The values of $a$ and $b$ are listed in Table \ref{tab:beta} together with the values of  the minimum number of repeat units, $n^\prime$, and the number of repeat units of the largest molecule in the series, $n_{max}$. Notice that $a$ quantifies the degree of scaling, and $b$ the hyperpolarizability in the limit of no repeat units.  In particular,
\begin{equation}\label{eq:a-beta}
a = \frac {\partial \beta_{int}} {\partial n},
\end{equation}
and
\begin{equation}\label{eq:b-beta}
b = \left. \beta_{int} \right|_{n=0}.
\end{equation}

\begin{table*}
\footnotesize
\caption{\small{The scaling parameters for the first hyperpolarizability molecular classes. The listed value of $E_{10}$ is the one for the molecule with the least number of repeat units in the class, denoted by $n^{\prime}$. $N$ is the number of effective electrons of the molecule with the least repeat units. $a$, $b$, $c$ and $d$ are the linear fitting parameters ($\beta_{int} = a n + b$ and $\beta_{int} = c \beta_{exp} + d$). $\beta_{int}^{max}$ is the value of the best intrinsic hyperpolarizability in the class, for the molecule that has $n_{max}$ repeat units. $\beta_{n=1}$ is the value of the absolute first hyperpolarizability for the molecule with $n=1$ repeat units in the series. $\beta_{SAT}$ is the predicted value of the absolute hyperpolarizablity at the saturation length, $FOM$ is the proposed figure of merit (Equation \ref{eq:FOM}) and $\Delta \beta_{exp}$ is the incremental addition to the absolute first hyperpolarizability per repeat unit (Equation \ref{eq:ratioac}).}}\label{tab:beta}
  \centering
  \begin{tabular}{|c | c | c | c | c | c | c | c | c | c | c | c | c | c |}
  \hline
  Class & $E_{10}$ & ~$n^{\prime}$~ & $N$ & $a$ & $b$ & $\beta_{int}^{max}$ & ~$n_{max}$ & $c$ (esu$^{-1}$) & $d$ & $\beta_{n=1}$ & $\beta_{SAT}$ & FOM & $\Delta \beta_{exp} $ \\
    & (eV) & & & $\times 10^{-3}$ & $\times 10^{-3}$ & $\times 10^{-3}$ & &$\times 10^{27}$ & $\times 10^{-3}$ & $ \times 10^{-30}$ & $ \times 10^{-28} $ & $ \times 10^{-30} $ & $ \times 10^{-30} $ \\
    & & & & & & & & & & esu & esu & esu & esu\\ \hline
  $\beta_1$\cite{cheng91.01} & 3.351 & 1  & 8 & -5 $\pm$ 1 & 21 $\pm$ 3 & 12 $\pm$ 2 & 2 & 2.0 $\pm$ 0.2 & -8 $\pm$ 2 & 4.5 & - & - & \\
  $\beta_2$\cite{cheng91.01} &  4.106 & 1 & 8 & -6 $\pm$ 1 & 23 $\pm$3 & 17 $\pm$ 3 &1 & -5.00 $\pm$ 0.02 & 31.0 $\pm$ 0.1 & 3.2 & - & - & \\
  $\beta_3$\cite{kuzyk98.01} &  4.429 & 0 & 8 & 2 $\pm$ 3 & 17 $\pm$5 & 24 $\pm$ 5 & 3 & 0.4 $\pm$ 0.5 & 16 $\pm$6 & 7 & - & - &\\
  $\beta_4$\cite{kuzyk98.01} & 3.229 & 0 & 8 & -1.0 $\pm$ 0.3 & 17.0 $\pm$ 0.7 & 16 $\pm$ 3 &1 & -0.20 $\pm$ 0.07& 18 $\pm$ 1 & 12.5 & - & - & \\
  $\beta_5$\cite{cheng91.01} &  3.647 & 1 & 16 & -1.0 $\pm$ 0.2 & 13.0 $\pm$ 0.4 & 12 $\pm$ 3 &1 & -0.30 $\pm$ 0.09 & 15$\pm$ 1 & 10.1 & - & - & \\
  $\beta_6$\cite{cheng91.01} &  3.483 & 1 & 16 & -1.0 $\pm$ 0.8 & 10.0 $\pm$ 0.2 & 9 $\pm$ 2 &1 & -0.40 $\pm$ 0.06 & 12 $\pm$ 6 & 8.8 & - & - & \\
  $\beta_7$\cite{cheng91.01} &  3.298 & 1  & 16 & -0.9 $\pm$ 0.3 & 13.0 $\pm$ 0.1 & 13 $\pm$ 3 &1 & -0.20 $\pm$ 0.08 & 15$\pm$ 2 & 14.9 & - & - &\\
  $\beta_8$\cite{cheng91.01} & 2.884 & 1 & 16 & -0.9 $\pm$ 0.4 & 12 $\pm$ 1 & 11 $\pm$ 2 &1 & -0.1 $\pm$ 0.1& 14 $\pm$ 4 & 21.2 & - & - & \\
  $\beta_9$\cite{marde94.02} & 2.768 & 0 & 12.5 & 22 $\pm$ 5 & 21 $\pm$ 12 & 90 $\pm$ 20 &3 & 0.080 $\pm$ 0.005& 25 $\pm$ 2 & 150 & 122 & 273 & 275 \\
  $\beta_{10}$\cite{marde94.02} & 2.375 & 0 & 12.5 & 19 $\pm$ 3 & 16 $\pm$ 6 & 77 $\pm$ 15 &3 & 0.06 $\pm$ 0.02 & 20 $\pm$ 1 & 186 & 163 & 315 & 317\\
  $\beta_{11}$\cite{marde94.02} & 2.460 & 0 & 12 & 14 $\pm$ 4 & 16 $\pm$ 9 & 55 $\pm$10 &3 & 0.09 $\pm$ 0.02 & 20 $\pm$ 6 & 100 & 109 & 155 & 156 \\
  $\beta_{12}$\cite{ortiz94.01} & 2.510 & 1 & 13.42 & -1 $\pm$ 1 & 14 $\pm$ 3 & 13 $\pm$ 3 &1 & -0.01 $\pm$ 0.05 & 13 $\pm$ 3 & 31 & - & - & \\
  $\beta_{13}$\cite{jayap99.01} & 2.510 & 1 & 4 & 90 $\pm$ 10 & 23 $\pm$ 30 & 378 $\pm$ 80 &4 & 0.30 $\pm$ 0.03 & 118 $\pm$ 17 & 66.8 & 29 & 271 & 300\\
  $\beta_{14}$\cite{jayap99.01} & 2.2828 & 1 & 4 & 65 $\pm$ 5 & 28 $\pm$ 11  & 220 $\pm$ 44 & 3 & 0.40 $\pm$ 0.09 & 84 $\pm$ 21 & 46.8 & 23 & 153 & 162 \\
  $\beta_{15}$\cite{meier05.01} & 4.106 & 1 & 8 & -9 $\pm$ 2 & 35 $\pm$ 3 & 27 $\pm$ 5 &1 & -3.00 $\pm$ 0.02 & 41.0 $\pm$ 0.1 & 5.1 & - & - & \\
  $\beta_{16}$\cite{meier05.01} & 2.799 & 0 & 12 & 27 $\pm$ 5 & 43 $\pm$ 13 & 160 $\pm$ 30 &4 & 0.090 $\pm$ 0.003 & 53 $\pm$ 2 & 195 & 105 & 297 & 300 \\
  $\beta_{17}$\cite{meier05.01} & --  & 0 & 10 & 22 $\pm$ 2 & 59 $\pm$ 8 & 170 $\pm$ 30 &5 & 0.070 $\pm$ 0.009 & 68 $\pm$ 9 & 195 & 133 & 311 & 314\\
  $\beta_{18}$\cite{jiang12.01} & 3.41 & 1 & 38 & -5 $\pm$ 3 & 63 $\pm$ 6 & 60 & 1 & -0.01 $\pm$ 0.01 & 61 $\pm$ 7 & 248 & - & - & \\
  \hline
\end{tabular}
\end{table*}

\begin{figure}
  \centering
  \includegraphics{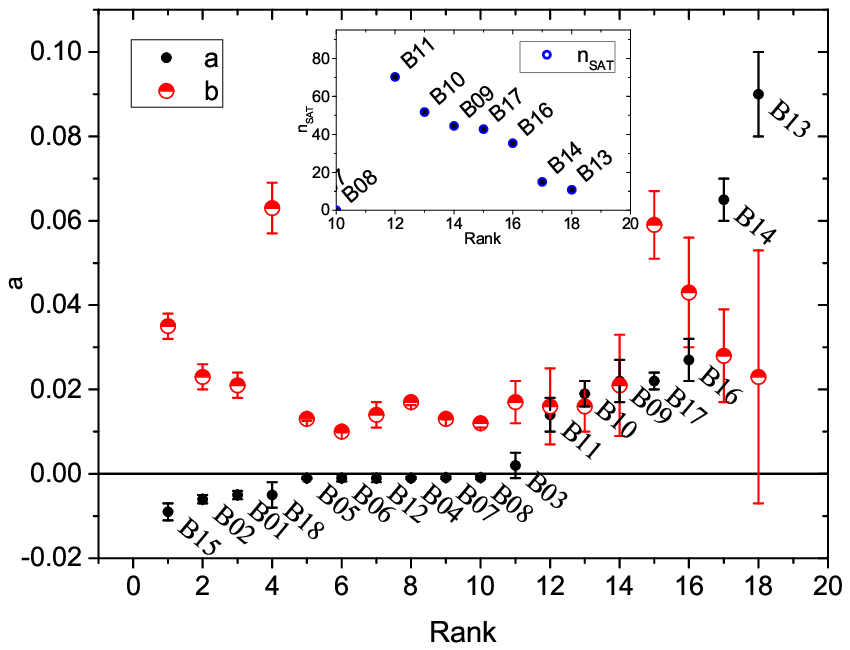}\\
  \caption{Plot of $a$ (The incremental addition to the intrinsic hyperpolarizability per repeat unit) and $b$ (the instrinsic hyperpolariability for no repeat units, i.e. $n=0$). On the horizontal axis, the classes are ranked based on the value of the highest value of $\beta_{int}$ in the class, $\beta_{int}^{max}$ (listed in Table \ref{tab:beta}). The inset shows the saturation figure of merit $n_{SAT}$ that gives the number of repeat units at which $\beta_{int}=1$ for the nominal and super-scaling classes.}\label{fig:a-and-b}
\end{figure}

Figure \ref{fig:a-and-b} plots $a$ and $b$ for all the molecular classes. On the horizontal axis, the classes are ranked based on the value of the highest value of $\beta_{int}$ in the class, $\beta_{int}^{max}$ (also listed in Table \ref{tab:beta}). Molecular classes with negative $a$ are in the sub-scaling group (B1, B2, B4, B5, B6, B7, B8, B12, B15 and B18). Molecular classes with $a \approx 0$ follow simple or nominal scaling (B03, B10, B11, B16 and B17). Molecular classes with positive $a$ follow super-scaling (B13 and B14). The value of $b$ for a class is its intrinsic hyperpolarizability in the zero-repeat unit limit.  Although class B18 has the largest value of $b$, increasing the length will not result in larger intrinsic nonlinearity, since B18 belongs to the sub-scaling group.

We argue that there are three criteria that define a molecular class that is a suitable paradigm for ultra-large nonlinear-optical response: (1) the absolute magnitude must be large; (2) scaling should be at least nominal so that longer homologues will have a hyperpolarizability that grows as a power law with length thus; (3) the ideal system should show no signs of saturation so that the large nonlinearity can be made bigger by simple scaling.

Since scaling should be at least nominal, a possible figure of merit is the number of repeat units required to attain an intrinsic hyperpolarizability $\beta_{int} = 1$, when the nonlinear response saturates ($\beta_{exp}=\beta_{max}$). The number of repeat units required to saturate the absolute hyperpolarizability, $n_{SAT}$, can be derived using Equation \ref{eq:linfitbetaint}:
\begin{equation}\label{eq:nSAT}
n_{SAT} = \frac {1-b} {a}.
\end{equation}
The better the molecular class, the smaller the value of $n_{SAT}$. The inset in Figure \ref{fig:a-and-b} shows the saturation length for the nominal and super-scaling classes (we have not included class B3 as $a$ could be negative due to the large error bars). Once again, classes B13 and B14 are the best.

\begin{figure}
  \centering
  \includegraphics[width=3.4in]{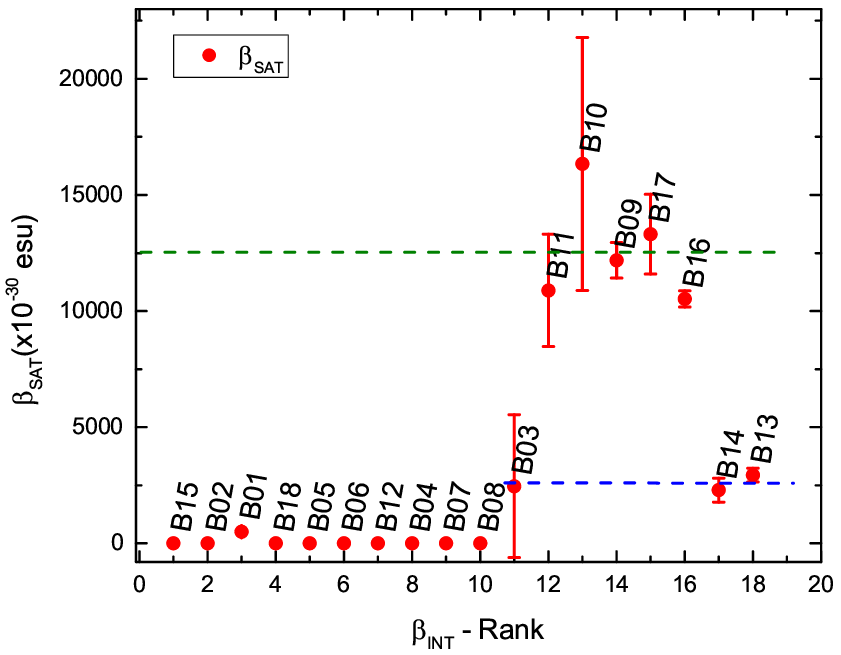}\\
  \caption{Plot of $\beta_{SAT}$, the expected absolute hyperpolarizability that would be achieved when the number of repeat units is saturated so that $\beta_{INT} = 1$. On the horizontal axis, the classes are ranked based on the highest value of $\beta_{int}$ in the class, $\beta_{int}^{max}$ (listed in Table \ref{tab:beta}).}\label{fig:BetaSAT1}
\end{figure}

However, the first criteria is that the absolute hyperpolarizability is large as the number of repeat units reaches saturation, where $\beta_{int} \rightarrow 1$. Figure \ref{fig:BetaSAT1} shows the predicted absolute hyperpolarizability when the number of repeat units saturates, $\beta_{SAT}$, as a function of the class rank. The class rank is determined from the value of $\beta_{int}^{max}$, the highest value of $\beta_{int}$ within the class. Although $n_{SAT}$ is not defined for classes that sub-scale, their saturation hyperpolarizability is well defined: since they sub-scale, the intrinsic hyperpolarizability will become zero as the number of repeat units is increased. The nominal and super-scaling classes are bunched into two groups. The first group is bunched around $\beta_{SAT} \approx 500 \times 10^{-30}$ esu (B3, B13 and B14), and the second group are classes with $\beta_{SAT} \approx 12500 \times 10^{-30}$ esu (B09, B10, B11, B16 and B17).

\begin{figure}
  \centering
  \includegraphics[width=3.4in]{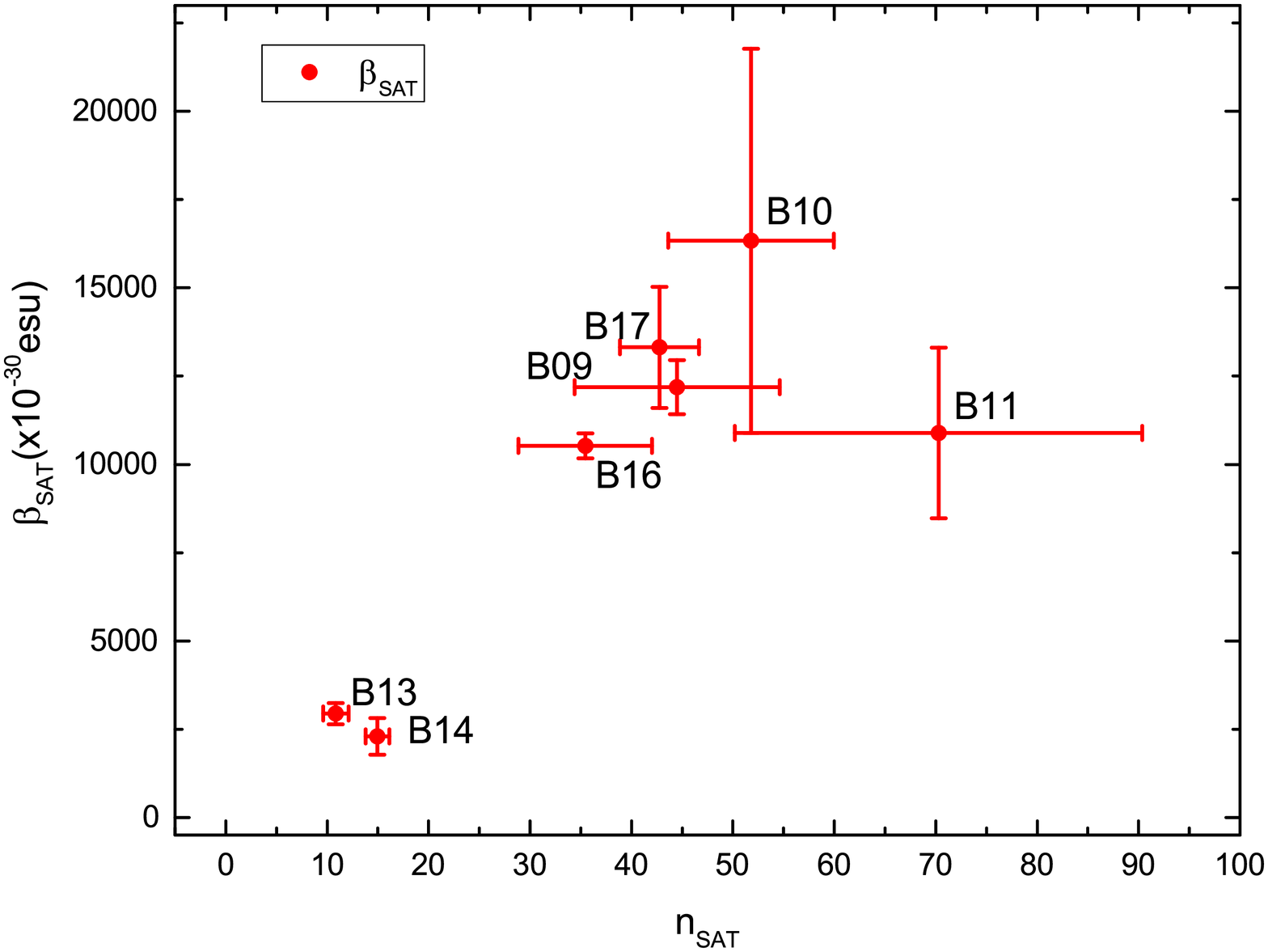}\\
  \caption{The expected absolute first hyperpolarizability that would be achieved when the class saturates ($\beta_{SAT}$), as a function of the number of repeat units needed to reach saturation. The expected absolute hyperpolarizability for classes that sub-scale (not shown in the plot) is zero.}\label{fig:BetaSAT}
\end{figure}

Figure \ref{fig:BetaSAT} shows the absolute hyperpolarizability when the number of repeat units saturates, $\beta_{SAT}$, as a function of $n_{SAT}$ for the nominal and super-scaling classes. The relationship between the two is approximately linear. Classes B13 and B14 (which were ranked best for their values of $n_{SAT}$) have the smallest values of $\beta_{SAT}$. The best absolute hyperpolarizability ($16300 \times 10^{-30}$ esu) would be achieved by class B10 with 50 repeat units.

It is highly unlikely that a molecule will retain its scaling properties once the length becomes long enough for environmental influences to interfere with conjugation. It is more reasonable to penalize the molecular classes with longest saturation length, $n_{SAT}$. Thus, a more appropriate figure of merit is given by:
\begin{equation}\label{eq:FOM}
FOM =  \frac{\beta_{SAT}} {n_{SAT}}.
\end{equation}

\begin{figure}
  \centering
  \includegraphics{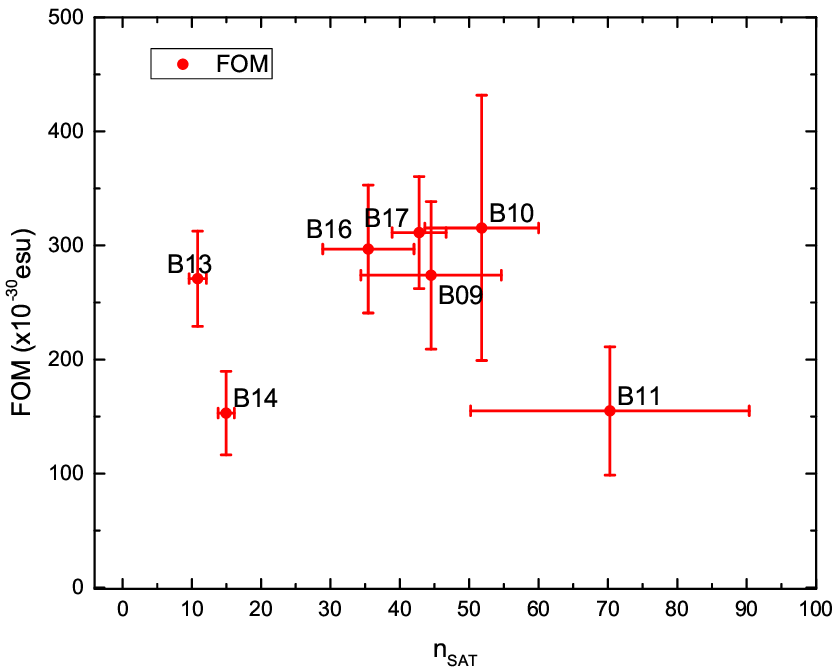}\\
  \caption{The figure of merit ($FOM$) defined as the ratio $\beta_{SAT} / n_{SAT}$ as a function of the number of repeat units needed to reach saturation.}\label{fig:FOM}
\end{figure}

Figure \ref{fig:FOM} shows a plot of the figure of merit ($FOM$) as a function of saturation length for nominal and super-scaling classes. All classes are found to share almost the same figure of merit within experimental uncertainties (between $150$ and $250 \times 10^{-30}$ esu), an illustration of the intimate relationship between these classes. This could be due to the fact that they all share a similar repeat unit based on the conjugation of carbon molecules.

The figure of merit (Equation \ref{eq:FOM}) is obtained from values extrapolated at the saturation length. Perhaps a more telling quantity is the incremental addition to the absolute hyperpolarizability with each repeat unit: $\Delta \beta_{exp} = \beta_{exp} (n) - \beta_{exp} (n-1)$, which applies to all lengths (it is the same for any value of $n$). Combining Equations \ref{eq:linfitintexp} and \ref{eq:linfitbetaint}, $\Delta \beta_{exp}$ can be expressed as a simple ratio of the fitting parameters $a$ and $c$:
\begin{equation}
\label{eq:ratioac}
\Delta \beta_{exp} = \frac{\Delta \beta_{int}}{c} = \frac{a \cdot(n)-a \cdot (n-1)}{c}=\frac{a}{c}.
\end{equation}

\begin{figure}
  \centering
  \includegraphics[width=3.4in]{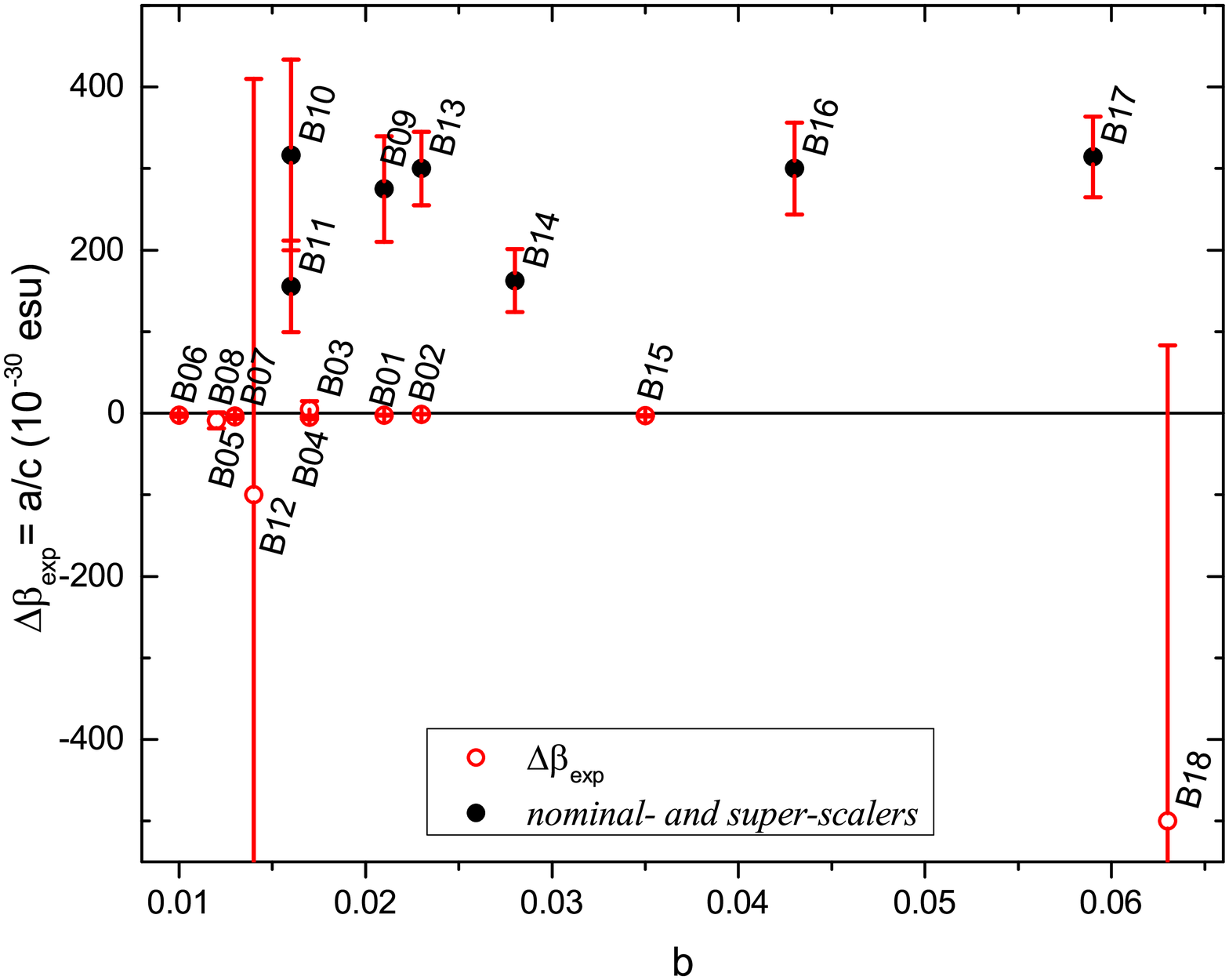}\\
  \caption{The incremental contribution to the absolute hyperpolarizability per repeat unit, $\Delta \beta_{exp}$, as a function of $b$ (the intrinsc hyperpolarizability in the limit of the base molecule, i.e. with $n=0$).}\label{fig:BetaPer}
\end{figure}

Figure \ref{fig:BetaPer} shows a plot of this incremental change as a function of $b$, the intrinsic hyperpolarizability in the limit of the base molecule, i.e. with $n=0$. Within experimental uncertainty, each of the super-scaling molecules (shown as black points with red error bars) have the same incremental contribution - not surprising given that each has the same conjugated bridge.  The last three columns in Table \ref{tab:beta} summarize all the various relevant quantities needed to assess nominal and super-scaling systems.

Having a polyene bridge, however, is not a guarantee that the incremental hyperpolarizability will be large. Many of the nominal scaling and sub-scaling molecules (those clustered near $a/c \approx 0$) have this type of bridge  but have much smaller incremental hyperpolarizability due to the effects of the end groups.

The best molecule with highest saturation hyperpolarizability, best figure of merit and largest incremental hyperpolarizability is B10. Interestingly, with the exception of molecule B16 - which has the lowest saturation hyperpolarizability of the group, all molecules (B09, B10, B11, and B17) with a polyene bridge and at least one of the end groups found in molecule B10 have the largest figure of merits. Thus, we conclude that the combination of a polyene-like bridge with one of the special end groups found in molecule B10 work synergistically to ensure that the hyperpolarizability super-scales {\em and} reaches the largest saturation hyperpolarizability.

B13 and B14 are the two molecules with the simplest structure, essentially a quasi one-dimensional wire that acts as a conduit for charge movement from one end of the molecule to the other during excitation.  These two molecules are anomalous in their large intrinsic hyperpolarizability, which is much larger than all others. They also super-scale and saturate at 10 to 15 repeat units, a number of units feasible for synthesis by chemists. In contrast, the next best molecule requires over 30 repeat units to reach saturation. The absolute hyperpolarizability, however, is not as large as what is attainable by the other structures. If more electrons could be added to this system while maintaining super scaling, this molecule class would have the potential for ultra-high hyperpolarizability.

\begin{figure}
  \centering
  \includegraphics[width=3.4in]{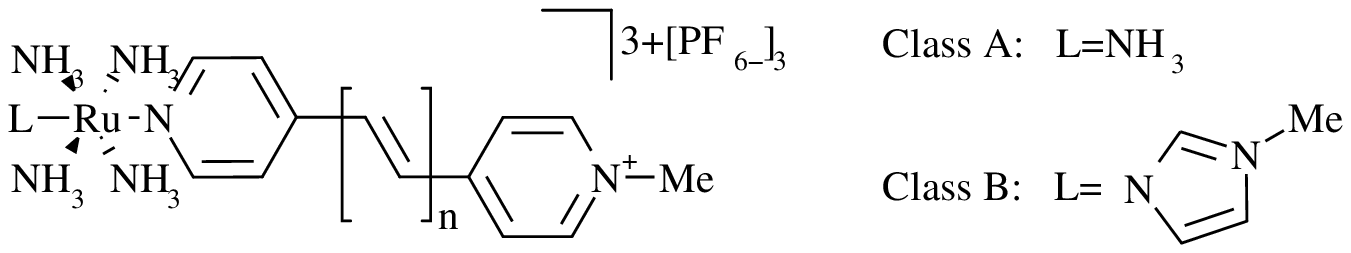}\\
  \caption{The two organometallic classes of Ruthenium(II) Ammine complexes considered in this work. Class $A$ consists of 4 homologues: $A1$ ($n=0$), $A2$ ($n=1$), $A3$ ($n=2$) and $A4$ ($n=3$). Class $B$ consists of 4 homologues: $B1$ ($n=0$), $B2$ ($n=1$), $B3$ ($n=2$) and $B4$ ($n=3$).}\label{fig:CoeSeriesAB}
\end{figure}

\begin{figure}
  \centering
  \includegraphics[width=3.4in]{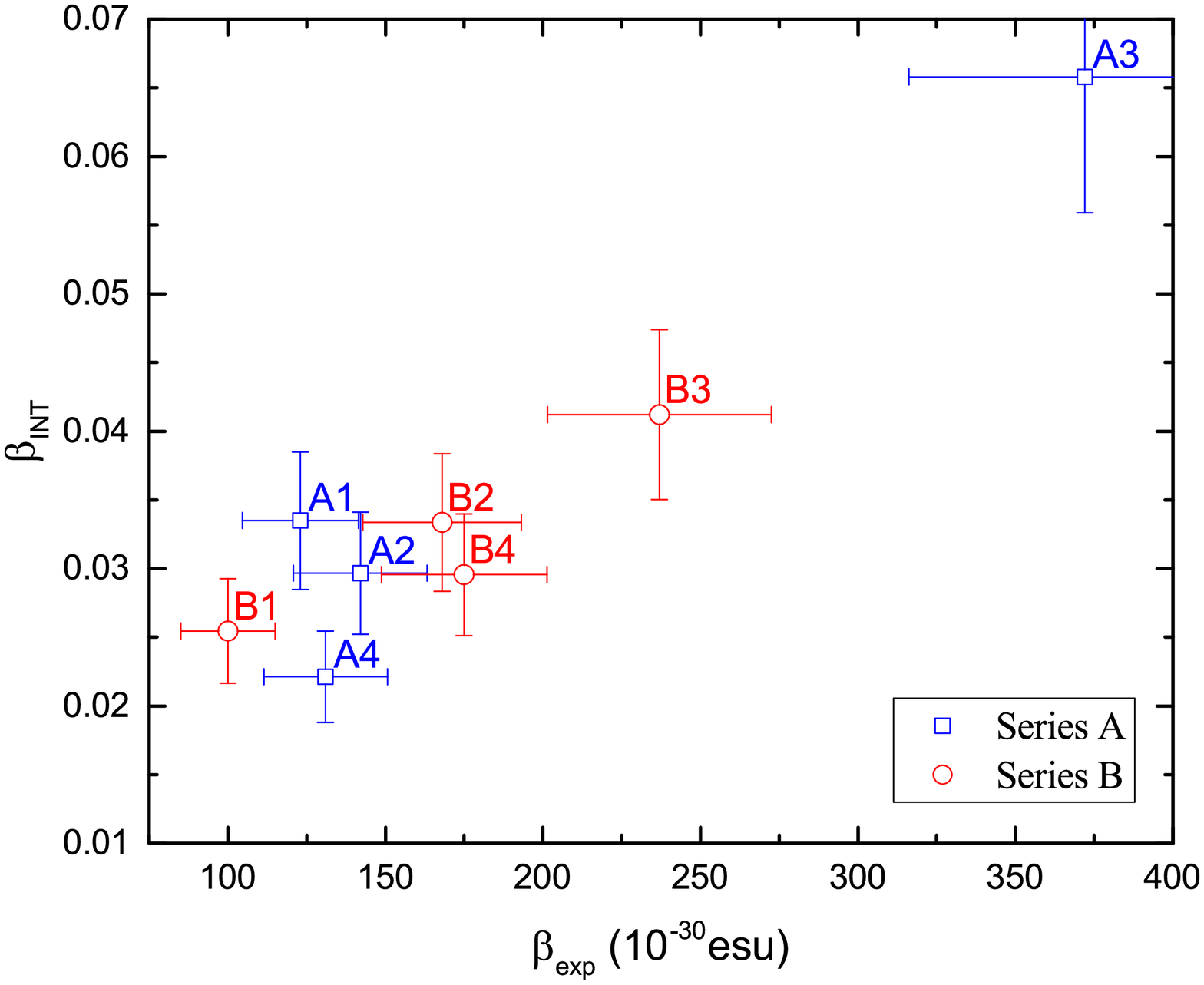}
  \caption{Plot of the intrinsic hyperpolarizability ($\beta_{int}$) as a function of the measured absolute hyperpolarizability ($\beta_{exp}$) for the two classes of organometallic Ruthenium(II) Ammine complexes.}
  \label{fig:CoeIntVsAbs}
\end{figure}

\begin{figure}
  \centering
  \includegraphics[width=3.4in]{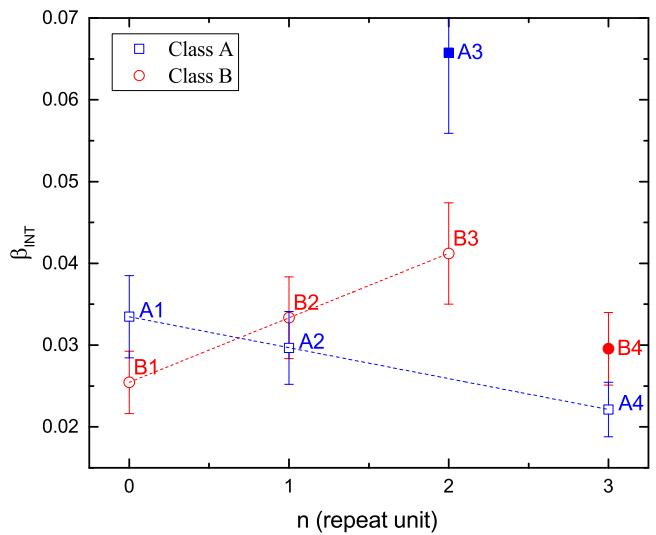}
  \caption{Plot of the intrinsic hyperpolarizability ($\beta_{int}$) as a function of the number of repeat units $n$ for the two classes of organometallic Ruthenium(II) Ammine complexes. The filled symbols are outliers that have not been included in the linear fitting $\beta_{int}(n)=a \cdot n + b$ (showed with dotted lines).}
  \label{fig:CoeIntVsN}
\end{figure}

Such organometallic structures warrant further investigation. Coe and coworkers have studied two different classes of Ruthenium(II) Ammine complexes shown in Figure \ref{fig:CoeSeriesAB}.\cite{coe03.02} Class $A$ is composed of 4 homologues: $A1$ ($n=0$), $A2$ ($n=1$), $A3$ ($n=2$) and $A4$ ($n=3$). Class $B$ is also composed of 4 homologues: $B1$ ($n=0$), $B2$ ($n=1$), $B3$ ($n=2$) and $B4$ ($n=3$). A plot of the intrinsic hyperpolarizability ($\beta_{int}$) as a function of the measured absolute hyperpolarizability ($\beta_{exp}$) for each class is shown in Figure \ref{fig:CoeIntVsAbs}. In general, a linear trend is not followed when all the data is considered. If we look at the dependence of the intrinsic hyperpolarizability on the number of repeat units, as shown in Figure \ref{fig:CoeIntVsN}, we can identify the outliers from each series (showed as filled symbols). The linear fit $\beta_{int}=a \cdot n + b$ (Equation \ref{eq:linfitbetaint}) for Class $A$ yields $a=(-3.78 \pm 0.01)\times 10^{-3}$ and $b=(3.347 \pm 0.002) \times 10^{-2}$. Therefore Class $A$ belongs to the sub-scaling group with $\beta_{SAT}=0$. The dependence of $\beta_{int}$ as a function of $\beta_{exp}$ is not linear and can not be fitted through Equation \ref{eq:linfitintexp}.

Class $B$ on the other hand follows nominal scaling. The linear fit $\beta_{int}=a \cdot n + b$ (Equation \ref{eq:linfitbetaint}) yields $a=(7.88\pm 0.02)\times 10^{-3}$ and $b=(2.545 \pm 0.002) \times 10^{-2}$. There is also a linear dependence between $\beta_{int}$ and $\beta_{exp}$ (Equation \ref{eq:linfitintexp}), which yields $c=(1.150 \pm 0.007)\times 10^{-4} (10^{30}\mbox{esu}^{-1})$ and $d=(1.3 \pm 0.6) \times 10^{-2}$. From these values, we can compute the saturation length (Equation \ref{eq:nSAT}) to be $n_{SAT} = (123.7 \pm 0.4) \approx 124$. This value is an order of magnitude larger than the values found for the other organometallic classes $B13$ ($n_{SAT} \approx 11$) and $B14$ ($n_{SAT} \approx 15$). It is also larger (more than double) than for any of the nominal and super-scaling classes. The absolute hyperpolarizability at the saturation length is predicted to be $\beta_{SAT}= 86 \times 10^{-28}$ esu, which is better than for classes $B13$ ($\beta_{SAT}= 29 \times 10^{-28}$ esu) and $B14$ ($\beta_{SAT}= 23 \times 10^{-28}$ esu), but lower than the other nominal and super-scaling classes (Figures \ref{fig:BetaSAT1} and \ref{fig:BetaSAT}). However, the figure of merit (Equation \ref{eq:FOM}) is $FOM=69.4 \pm 0.2) \times 10^{-30}$ esu, which is lower than the ones from the rest of the nominal and super-scaling classes (Figure \ref{fig:FOM}). Finally, the incremental addition to the absolute hyperpolarizability per repeat unit (Equation \ref{eq:ratioac}) is given by $\Delta \beta = (68.5 \pm 0.6) \times 10^{-30}$ esu, which is again, the lowest when compared wit the other nominal and super-scaling groups. This explains why so many more units are needed in order to reach the saturation of the first hyperpolarizability.

The theory of the quantum limits can be used to study the nature of the nonlinear response. Using the three-level ansatz, the diagonal component of the absolute first hyperpolarizability in the off-resonance regime is simplifies to
\begin{equation}
\beta(E,X) = \beta_{max} \cdot f(E) \cdot G(X),
\label{eq:betaEX}
\end{equation}
where $\beta_{max}$ is the quantum limit (Equation \ref{eq:betaMax}), and the functions $f(E)$ and $G(X)$ are defined as
\begin{equation}
f(E)=\frac{1}{2} (1-E)^{3/2} (2+3E+2E^{2})
\label{eq:fE}
\end{equation}
and
\begin{equation}
G(X)=\sqrt[4]{3} X \sqrt{\frac{3}{2}(1-X^{4})}.
\label{eq:GX}
\end{equation}
These functions are expressed in terms of the dimensionless parameters $E$ and $X$ defined as:
\begin{equation}
E=\frac{E_{10}}{E_{20}}
\label{eq:Edef}
\end{equation}
and
\begin{equation}
X= \frac{|x_{01}|}{\sqrt{\frac{\hbar^{2} N}{2 m E_{10}}}},
\label{eq:Xdef}
\end{equation}
where $E_{n0}$ is the energy difference between the excited state $|n\rangle $ and the ground state $|0\rangle$, while $|x_{01}|$ represents the magnitude of the transition dipole moment between states $|0\rangle$ and $|1\rangle$. Notice that $E$ and $X$ can only range between $0$ and $1$. $f(E)$ and $G(X)$ are optimized separately, achieving maximum values when $E \rightarrow 0$ and $ X = \sqrt[4]{1/3}$, with $f_{max} = f(0)=1$ and $G_{max}=G(\sqrt[4]{1/3})=1$.

The values of $E_{10}$ and $|x_{01}|$ can be obtained experimentally through linear absorption spectroscopy. This allows one to compute $X$ using Equation \ref{eq:Xdef} followed by substitution into Equation \ref{eq:GX} to obtain $G(X)$. Then $f(E)$ is computed according to\cite{tripa04.01, tripa06.01, perez07.02}
\begin{equation}
f(E)=\frac{\beta_{exp}}{\beta_{max} \cdot G(X)} = \frac{\beta_{int}}{G(X)}.
\label{eq:fEfromExp}
\end{equation}
Once $f(E)$ is determined from the experimental data, we can invert Equation \ref{eq:fE} to find the predicted ratio of energies $E$ and solve for $E_{20}$. This value can be compared with the value of $E_{20}$ gathered through linear absorption spectroscopy, in order to determine how well the response is described by only three states. It is important to note that in a system with many levels, the energy function $f(E)$ and the ratio of energies $E$ act as a proxy for the energy-level spacing of the molecule: the closer $f(E)$ is to $1$, the better the arrangement of energies in the molecule.

\begin{table*}
\footnotesize
\caption{\small{The relevant parameters for the quantum limits analysis applied to the two classes of organometallic Ruthenium(II) Ammine complexes (Class A and Class B).}}\label{tab:fEGX}
\centering
\begin{tabular}{|c|c|c|c|c|c|c|c|c|c|}
\hline
Compound & $E_{10}$ & n & $\beta_{exp}$ & $\beta_{int}$ & $X$ & $G(X)$ & $f(E)$ & Predicted & Experimental \\
& (eV) & (repeat units) & ($10^{-30}$ esu) & ($10^{-2}$) & & & & $E_{20}$ (eV) & $E_{20}$ (eV) \\
\hline
A1 & 2.1 & 0 & 123 $\pm$ 18 & 3.3 $\pm$ 0.5 & 0.268 $\pm$ 0.004 & 0.431 $\pm$ 0.007 & 0.07 $\pm$ 0.01 & 2.29 & 4.6 \\
A2 & 2.08 & 1 & 142 $\pm$ 21 & 3.0 $\pm$ 0.4 & 0.255 $\pm$ 0.004 & 0.441 $\pm$ 0.007 & 0.07 $\pm$ 0.01 & 2.26 & 3.9 \\
A3 & 2.1 & 2 & 372 $\pm$ 56 & 6.6 $\pm$ 1.0 & 0.263 $\pm$ 0.004 & 0.423 $\pm$ 0.006 & 0.16 $\pm$ 0.03 & 2.46 & 3.5 \\
A4 & 2.18 & 3 & 131 $\pm$ 20 & 2.2 $\pm$ 0.3 & 0.249 $\pm$ 0.004 & 0.401 $\pm$ 0.006 & 0.06 $\pm$ 0.01 & 2.33 & 3.2 \\
\hline
\hline
B1 & 2.06 & 0 & 100 $\pm$ 15 & 2.5 $\pm$ 0.4 & 0.274 $\pm$ 0.004 & 0.441 $\pm$ 0.007 & 0.06 $\pm$ 0.01 & 2.21 & 4.6 \\
B2 & 2.05 & 1 & 168 $\pm$ 25 & 3.3 $\pm$ 0.5 & 0.262 $\pm$ 0.004 & 0.421 $\pm$ 0.007 & 0.08 $\pm$ 0.01 & 2.34 & 3.9 \\
B3 & 2.09 & 2 & 237 $\pm$ 36 & 4.1 $\pm$ 0.6 & 0.268 $\pm$ 0.004 & 0.459 $\pm$ 0.006 & 0.09 $\pm$ 0.01 & 2.29 & 3.5 \\
B4 & 2.18 & 3 & 175 $\pm$ 26 & 3.0 $\pm$ 0.4 & 0.290 $\pm$ 0.004 & 0.466 $\pm$ 0.006 & 0.06 $\pm$ 0.01 & 2.35 & 3.1 \\
\hline
\end{tabular}
\end{table*}

\begin{figure}
  \centering
  \includegraphics[width=3.4in]{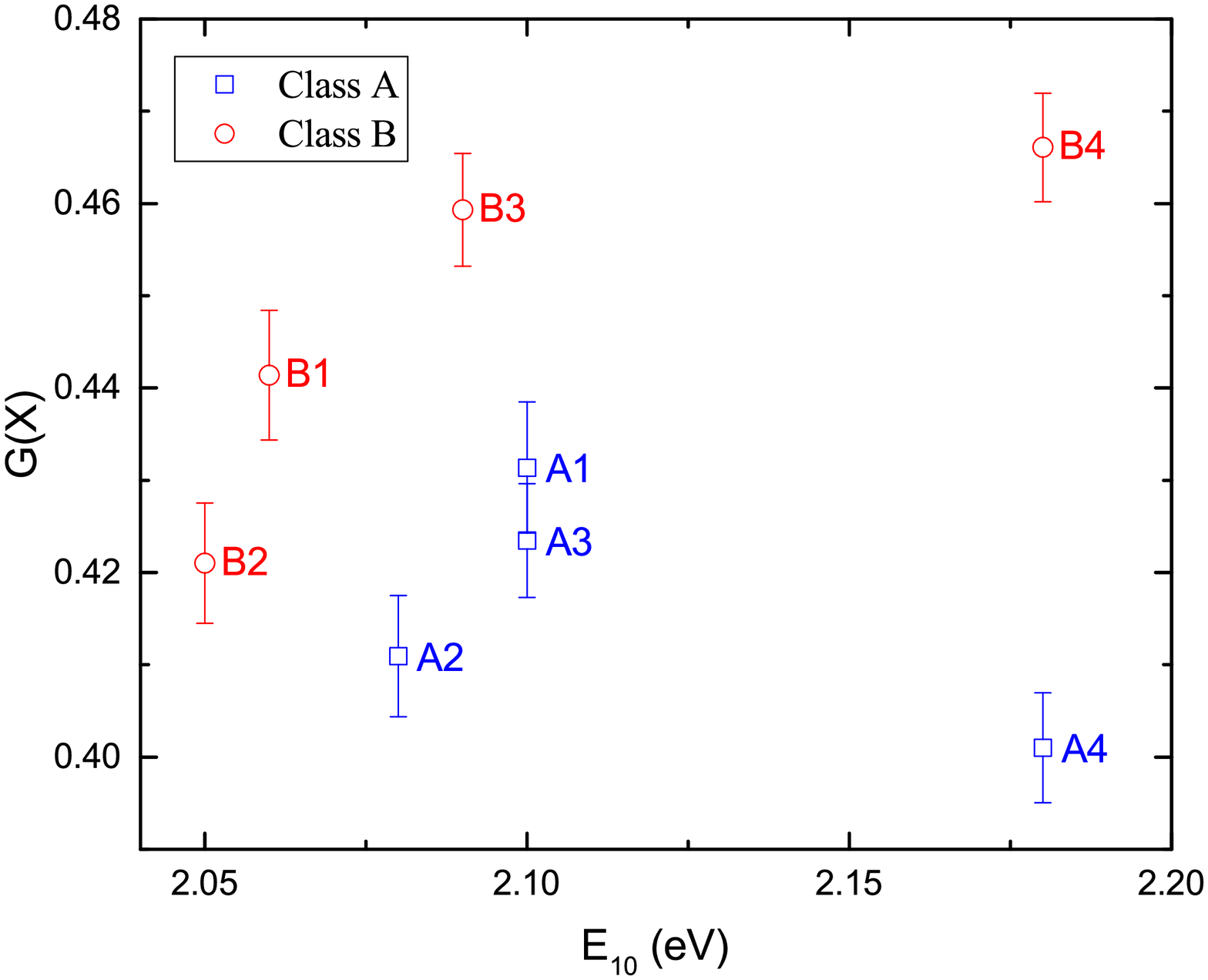}
  \caption{Plot of $G(X)$ as a function of $E_{10}$ for each compound of the two classes of organometallic Ruthenium(II) Ammine complexes (Class A and Class B).}
  \label{fig:GX}
\end{figure}

\begin{figure}
  \centering
  \includegraphics[width=3.4in]{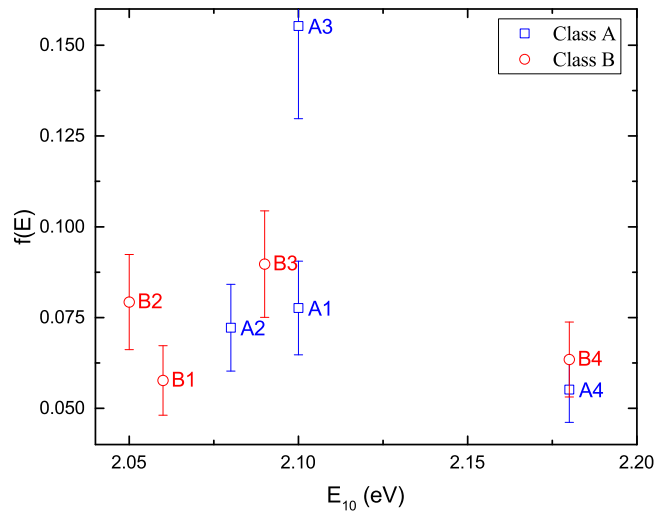}
  \caption{Plot of $f(E)$ as a function of $E_{10}$ for each compound of the two classes of organometallic Ruthenium(II) Ammine complexes (Class A and Class B).}
  \label{fig:fE}
\end{figure}

Detailed linear absorption spectroscopy was performed for the two collections of organometallic molecules,\citep{coe03.02, coe1999tuning, coe1997large, coe1998enhancement, coe2004syntheses} so we can apply the quantum limits analysis to Classes $A$ and $B$. The values of $X$, $G(X)$, $f(E)$ and other relevant parameters for the two classes are summarized in Table \ref{tab:fEGX}. Figure \ref{fig:GX} plots $G(X)$ as a function of $E_{10}$ for each compound of the two classes, and Figure \ref{fig:fE} plots $f(E)$ as a function of $E_{10}$.

A general trend for all the other molecular classes is that the value of $E_{10}$ decreases as the number of repeat units increases.\cite{kuzyk13.01} This trend is not followed by class A and class B. In fact, the values of $E_{10}$ are very similar for all the compounds, within the range of $2.1 (\pm 0.1)$ eV. The values of $X$ and $G(X)$ are similar and fall within the typical range for this type of molecule ($0.2 \leq G(X) \leq 06 $).\citep{tripa04.01, tripa06.01, perez07.02} With regards to the behavior of $f(E)$, for class A the values are all similar with the exception of the outlier $A3$.

For class B, $f(E)$ increases slightly with the number of repeat units (if we ignore the outlier $B4$).  In general, the values of $f(E)$ for these types of molecules are found between $f(E) \approx 0.01$ and $f(E) \approx 0.10$, while the values for these molecules are all near $f(E) \approx 0.07$. Thus, the quantum limits analysis suggests that the molecules of these two classes behave similarly. We notice that $\sqrt{\frac{\hbar^{2}N}{2m E_{10}}}$ increases as the number of repeat units increases, so the fact that $X$ remains approximately constant implies that the experimental value of $|\mu_{10}|$ does also increase. Although the variations of $f(E)$ are also very small, we can conclude that changes in the length of these molecules result in more or less favorable energy distributions, while keeping a similar transition dipole moment distribution. However, the predicted values of $E_{20}$ are far from the experimental values, and hence, the experimental data indicate that more than tree levels contribute to the response, so the validity of the three-level analysis is limited.

\section{Approach to analyzing molecular series}

The importance of the present work is no sot much in the results that we have presented, which serve as an example, but in the protocols that we define for a methodology that identifies the promising series of molecules for further study and optimization for scale-up. Based on the examples above, we propose the following approach.

The goal is to find the ideal unit that can be scaled up by linking the units together. The simplest units are ones that connect to form linear changes, but others are possible, including the formation of dendrimers, space filling structures, or any other novel shapes. The units can be stand alone; used to link two ends together, thus having the type of ends as an additional degree of freedom; or can be formed into fractal-like dendrimer units with multiple external units and joints.

The process proceeds as follows:

\begin{enumerate}

\item \label{step-Type}
Identify a structure type that includes repeat units and end/exterior units that is expected to show promise based on semi-empirical calculations or intuition

\item \label{step-ChooseEnds}
Choose end/exterior units and keep them fixed and synthesize a series of structure of varying length between 1 and 4 or 5 repeat units.

\item
Measure the linear absorption spectrum for each to determine $\beta_{max}$, and then measure $\beta$ as a function of the number of repeat units to determine $\beta_{int}$.

\item
From a linear fit of $\beta_{int}$ versus $\beta_{exp}$ and $\beta_{int}$ versus $n$, determine $n_{SAT}$, $\beta_{SAT}$ and the figure of merit.

\item
If the figure of merit for $\beta$ is larger than $2 \times 10^{-28} esu$ then make structures with a large number of repeat units. Otherwise, go to Step \#\ref{step-ChooseEnds}.

\item
If the scaling law breaks down for longer units, start again at Step \#\ref{step-Type}

\end{enumerate}

This procedure identifies useful paradigms that have the potential for ultra-large second-order nonlinear-optical response. Since making lager molecules is a more involved process, the proposed methodology identifies a series that is worth the effort.

\section{Conclusion}

Making a direct comparison of the nonlinear-optical response of two molecules is problematic because they may be of differing sizes, so differences may be due solely to simple scaling and not to the intrinsic nonlinear response of the molecule. The size of a molecule is not well defined from the quantum perspective because molecules do not have sharp boundaries. However, the difference in energy between the first excited state and the ground state, $E_{10}$, and the effective number of electrons, $N$, define a size, which is embodied in the fundamental limit of the second-order nonlinear response, $\beta_{max}$, a function of only $N$ and $E_{10}$. Dividing the nonlinear response by the fundamental limit defines the intrinsic response, which is a scale invariant property that can be used to compare molecules of disparately different sizes. Indeed, the range of the intrinsic nonlinearity is much smaller than the absolute nonlinearities because much of the difference is due to size effects.

Using the idea of scale invariance, we have introduced a protocol and defined a figure of merit that can be used to compare a series of molecules that differ mostly in their lengths. This protocol can be used to identify new paradigms that are scalable; that is, longer versions of the molecule return a nonlinearity that is far larger than one would attain if it were due only to the increased length.

We have shown how this method can be used to analyze which material classes are the most promising for second-order nonlinear optical applications. We find that that a simple polynene bridge with complex cyclic end groups as embodied in class B10 is the ideal structure. This molecule, and minor modifications of it, shows promise for ultra-large nonlinearity if it can be scaled up.

More importantly, our work uses a review of the literature to illustrate a new approach for identifying better molecular classes. Using this type of well defined procedure may be required to make the next big leap in the design of new molecules.

\section{Funding Information}

We acknowledge the National Science Foundation(ECCS-1128076) for generously supporting this work.

\bibliography{\bibs}

\end{document}